\documentclass{article}
\usepackage{arxiv}

\usepackage[utf8]{inputenc} % allow utf-8 input
\usepackage[T1]{fontenc}    % use 8-bit T1 fonts
\usepackage{hyperref}       % hyperlinks
\usepackage{adjustbox}
\usepackage{url}            % simple URL typesetting
\usepackage{booktabs}       % professional-quality tables
\usepackage{framed,multirow}
\usepackage{amsfonts,amssymb,amsmath}       % blackboard math symbols
\usepackage{nicefrac}       % compact symbols for 1/2, etc.
\usepackage{microtype}      % microtypography
\usepackage[bottom]{footmisc}
\usepackage{lipsum}
\usepackage[numbers]{natbib}
\usepackage{doi}
\usepackage{fancyhdr}       % header
\usepackage{graphicx}       % graphics
\usepackage{float}
\usepackage{wrapfig}
\usepackage{soul}
\usepackage{bm}
\usepackage[nameinlink]{cleveref}
\usepackage{textcomp}
\usepackage{latexsym}
\usepackage{colortbl}
\usepackage{letltxmacro}
\usepackage{xcolor}
\usepackage{subcaption}
\usepackage{booktabs}
\usepackage{array}
\usepackage{makecell}
\usepackage{graphicx}
\usepackage{rotating}
\usepackage{pdflscape}
\usepackage{romannum}

\hypersetup{
    colorlinks=true,
    linkcolor=teal,
    filecolor=cyan,
    urlcolor=magenta,
    allcolors=blue
}

\newcommand{\mcal}[1]{\mathcal{#1}}

\def\vs{\emph{vs}\onedot}

\definecolor{darkgreen}{rgb}{0,0.6,0}

\def\eqref#1{equation~\ref{#1}}

\def\1{\bm{1}}

\def\rvw{{\mathbf{w}}}
\def\rvx{{\mathbf{x}}}
\def\rvy{{\mathbf{y}}}

\def\vtheta{{\bm{\theta}}}

\def\va{{\bm{a}}}

\def\vs{{\bm{s}}}

\def\vu{{\bm{u}}}

\def\vz{{\bm{z}}}

\def\mA{{\bm{A}}}

\def\mD{{\bm{D}}}

\def\mI{{\bm{I}}}

\def\mT{{\bm{T}}}

\DeclareMathAlphabet{\mathsfit}{\encodingdefault}{\sfdefault}{m}{sl}
\SetMathAlphabet{\mathsfit}{bold}{\encodingdefault}{\sfdefault}{bx}{n}

\DeclareMathOperator*{\argmin}{arg\,min}

\newcommand{\ud}{\mathop{}\!\mathrm{d}}

\newcommand{\citea}[1]{{\hypersetup{hidelinks}\citeauthor{#1}}
 \cite{#1}}

\title{%Review of Physics-inspired Generative AI for Medical Imaging
Physics-inspired Generative Models in Medical Imaging: A Review}

\usepackage{authblk}

\author[1,2]{Dennis~Hein\hspace{0.75pt}\thanks{Equal contributions}}
\author[3]{Afshin~Bozorgpour$^{*}$}
\author[3,4]{Dorit~Merhof\hspace{0.75pt}\thanks{Corresponding authors}}
\author[5]{Ge~Wang$^{\dag}$}

\affil[1]{Department of Physics, KTH Royal Institute of Technology, Stockholm, Sweden}
\affil[2]{MedTechLabs, Karolinska University Hospital, Stockholm, Sweden}
\affil[3]{Faculty of Informatics and Data Science, University of Regensburg, Regensburg, Germany}
\affil[4]{Fraunhofer Institute for Digital Medicine MEVIS, Bremen, Germany}
\affil[5]{Department of Biomedical Engineering, School of Engineering, Biomedical Imaging Center, Center for Biotechnology and Interdisciplinary Studies, Rensselaer Polytechnic Institute, Troy, NY, USA}

\renewcommand{\headeright}{}
\renewcommand{\undertitle}{}
\renewcommand{\shorttitle}{%Review of Physics-inspired Generative AI for Medical Imaging
Physics-inspired Generative Models in Medical Imaging: A Review}

\hypersetup{
    pdftitle={%Review of Physics-inspired Generative AI for Medical Imaging
Physics-inspired Generative Models in Medical Imaging: A Review},
    pdfsubject={q-bio.NC, q-bio.QM},
    pdfauthor={Dennis~Hein, Afshin~Bozorgpour, Dorit~Merhof, Ge~Wang},
    pdfkeywords={Physics-inspired generative models,
Bayesian Theorem,
Diffusion Model,
Poisson Flow Generative Model,
PFGM++, 
Consistency Model,
Medical Imaging,
Image Reconstruction,
Image Analysis,
Image/Data Synthesis},
}

\begin{document}
    \maketitle
    
    % abstract
    \begin{abstract}
Physics-inspired Generative Models (GMs), in particular Diffusion Models (DMs) and Poisson Flow Models (PFMs), enhance Bayesian methods and promise great utility in medical imaging. 
This review examines the transformative role of such generative methods. First, a variety of physics-inspired GMs, including Denoising Diffusion Probabilistic Models (DDPMs), Score-based Diffusion Models (SDMs), and Poisson Flow Generative Models (PFGMs and PFGM++), are revisited, with an emphasis on their accuracy, robustness as well as acceleration. Then, major applications of physics-inspired GMs in medical imaging are presented, comprising image reconstruction, image generation, and image analysis. Finally, future research directions are brainstormed, including unification of physics-inspired GMs, integration with Vision-Language Models (VLMs), and potential novel applications of GMs. Since the development of generative methods has been rapid, this review will hopefully give peers and learners a timely snapshot of this new family of physics-driven generative models and help capitalize their enormous potential for medical imaging.

\end{abstract}

%old:
%Physics-inspired score-matching models, especially diffusion models and Poisson flow generative models, empower Bayesian methods and promise major utilities for medical imaging. This review examines the transformative role of such generative artificial intelligence (AI) methods. The progression of generative AI methods already generated exciting medical imaging results and promises an enormous potential in tomographic reconstruction, image analysis, and content generation. Future research directions are also discussed.
    
    % keywords can be removed
    
    % ================= SECTIONS ==================
    
    % introduction
    
\section{INTRODUCTION}
The Bayesian theorem, which is fundamental to many fields, is expressed as $P(A|B)=P(B|A)P(A)/P(B)$, where $P(A|B)$ is the posterior probability that event $A$ will occur given that $B$ is true, and $P(B|A)$ is the likelihood defined as the probability that event $B$ will occur given that $A$ is true. Many medical imaging problems are actually special cases of Bayesian inference. The Bayesian formula embodies the essence of rational decision-making under uncertainty, provides a rigorous framework to update our belief based on new evidence, and has led to numerous successes in major applications and even in philosophical arguments about the nature of knowledge and learning. By its quantifiable, coherent and powerful nature, the Bayesian approach elegantly serves as a cornerstone of modern science and technology.

In tomographic image reconstruction, $A$ is an image to be reconstructed from tomographic data $B$. For many years, maximum a posteriori (MAP) image reconstruction has been the holy grail to reconstruct an optimal image that is most probable given the observed data and prior knowledge about the underlying images. By maximizing the posterior probability, MAP reconstruction balances the fidelity to the observed data with regularization imposed by the prior. However, it is well known that the main challenge in achieving the MAP reconstruction (and many other tasks) is often the lack of a good model for the distribution of $A$. Assuming $P(A)$ to be constant, the maximum likelihood estimation was widely used as a surrogate of the MAP estimation. To approximate $P(A)$, roughness penalty, sparsity constraint, and low-rank prior were subsequently introduced, but these classic regulators are quite simplistic, commonly expressed mathematically in no more than one line. Indeed, an image content $A$ in the real world can have rather sophisticated and diverse features, being typically on a low-dimensional manifold.  Until approximately one decade ago, modelling a highly complex data distribution has been an intractable task.

Currently, the landscape of artificial intelligence (AI) and machine learning, especially deep learning, is rapidly evolving, with generative AI at the forefront of innovation. Contemporary generative AI methods have revolutionized the field of AI and machine learning by enabling the creation of new, synthetic instances of data that are practically indistinguishable from real ones. These techniques have widespread applications, with Sora and GPT-4o as two amazing examples. 

In retrospect, Variational Autoencoders (VAEs) and Generative Adversarial Networks (GANs) were among the first to pave the way. VAEs work by encoding input data into a lower-dimensional space and then synthesize samples in the original space, obeying the distribution of the original data. GANs, on the other hand, consist of two competing networks: a generator that creates data and a discriminator that separates generated and real data. The two networks interact to define the adversarial process that improves the quality of the generated data so that they are indistinguishable from the original data.  More recently, physics-inspired Generative Models (GMs) have become the state of the art, with the Diffusion Models (DMs) and Poisson Flow Generative Models (PGFMs) being the most popular ones. They construct an image or sample by starting from a random distribution and refining it towards the distribution of the target data through a sampling process in a number of steps (from thousands in the original setting to one in the consistency model). These physics-inspired GMs can generate realistic, diverse, and complex data in a principled yet universal way, already demonstrating a huge potential for medical imaging. Notably, as we will see in this review, VAEs are making a silent comeback as a key construct playing a pivotal role in the state-of-the-art generative methods.

Given the importance and momentum of generative AI research, there are a good number of review papers already under this theme. We used the Scopus rule in the Article Title field “Survey OR Review OR Overview” and obtained 113 documents. After examination, not all these reviews are relevant to generative AI or of the same quality. Furthermore, these reviews are not up-to-date, given the rapid evolution of the field. 
Also, we have an excellent GitHub repository ``Awesome-Diffusion-Models", \url{https://github.com/diff-usion/Awesome-Diffusion-Models}, which is a comprehensive collection of curated documents dedicated to the study and application of diffusion-type models, including introductory posts, papers, videos, lectures, tutorials, and Jupyter notebooks on various aspects of such models. Based on our Scopus$^{\text{TM}}$ search results and this repository, in the following we comment on the most recent and most comprehensive reviews that are closely related to medical imaging. 

An overview of DMs was first posted in November 2022, revised in June 2023, and formally published in Aug 2023 in Medical Image Analysis~\cite{kazerouni2023}. It cited 192 key papers and provides an in-depth coverage of the theoretical foundations of DMs, their practical applications in medical imaging using the taxonomy built on nine sub-fields such as image translation, reconstruction, and anomaly detection, as well as the challenges and future directions. The survey categorized DMs based on their applications, imaging modality, organ of interest, and algorithms, offering a multi-perspective categorization and highlighting the importance of these models in generating synthetic medical images. Such realistically simulated samples can alleviate data scarcity and improve network training for superior performance. 

Another interesting review published in August 2023 analyzed 80 papers on image generation using DMs~\cite{zhang2023}. The difficulties were examined in generating images with multiple objects, creating images of rare/novel objects and improving quality of generated images. The paper organized the discussion around these challenges and presented various methods to address them, such as leveraging layout information, refining attention maps, and employing retrieval-based methods. It also touched upon the importance of text encoders in improving image quality and explored the use of mixtures of experts and reinforcement learning to enhance outcomes. 

Another survey over 160 papers was posted in 2022 and published in IEEE Transactions on Pattern Analysis and Machine Intelligence (TPAMI) in September 2023~\cite{croitoru2023}. It covered DMs in the field of computer vision and addressed the three schemes: Denoising Diffusion Probabilistic Models (DDPMs), Noise Conditioned Score Networks (NCSNs), and Score-based Diffusion Models (SDMs) via stochastic differential equations (SDEs). The paper also discussed the relationship between DMs and other GMs, presenting a categorization of DMs applied in computer vision and identifying research directions, particularly the need to improve sampling/inference efficiency without compromising the quality of generated samples. In yet another comprehensive review of 258 papers on DMs in various domains, the core aspects were presented of the formulation, algorithms, and applications~\cite{cao2023}. Through gradual refinement of the review from September 2022 (version 1) to December 2023 (version 10), this review highlighted methods for generating high-fidelity samples across different modalities and discussed strategies for sampling acceleration, diffusion process design, and likelihood optimization. The article was concluded with discussions on the current limitations and future advancements. 

While each review had different methodological emphases and covered various spectra of applications, they have collectively underscored the significance of DMs across different fields, with the common core challenge to optimize the image quality and sampling efficiency individually or simultaneously.

While the above reviews/surveys are excellent and complementary, the rationale for us to write this new review is still compelling. Recently, especially over the past year, the emergence of PFGMs~\cite{xu2022, xu2023} represents a significant leap forward in the field of generative AI, modeling complex data distributions and producing high-fidelity samples with the state-of-the-art performance. Like DMs, Poisson flow or other physics-/mathematics-inspired GMs can be cast in the score-matching format. Unlike DMs, which draw their inspiration from non-equilibrium thermodynamics, PFGM are grounded in the principle of electrostatics, bringing new theoretical insights and empowering technical capabilities. This foundational shift offers a distinct advantage in terms of precision and efficiency. Poisson equations allow for a more direct manipulation of the data distribution, enabling a better controlled and potentially faster convergence to the target distribution. At the same time, impressive new results have been reported on the classic DMs such as consistency models. Therefore, the benefits to thoroughly summarize and analyze these advancements through a new review article are clear; mainly, to clarify the new thinking and mechanisms, capitalize the theoretical and technical potentials, and facilitate real-world applications of contemporary generative AI. 

\begin{figure}
    \centering
    \includegraphics[width=\linewidth]{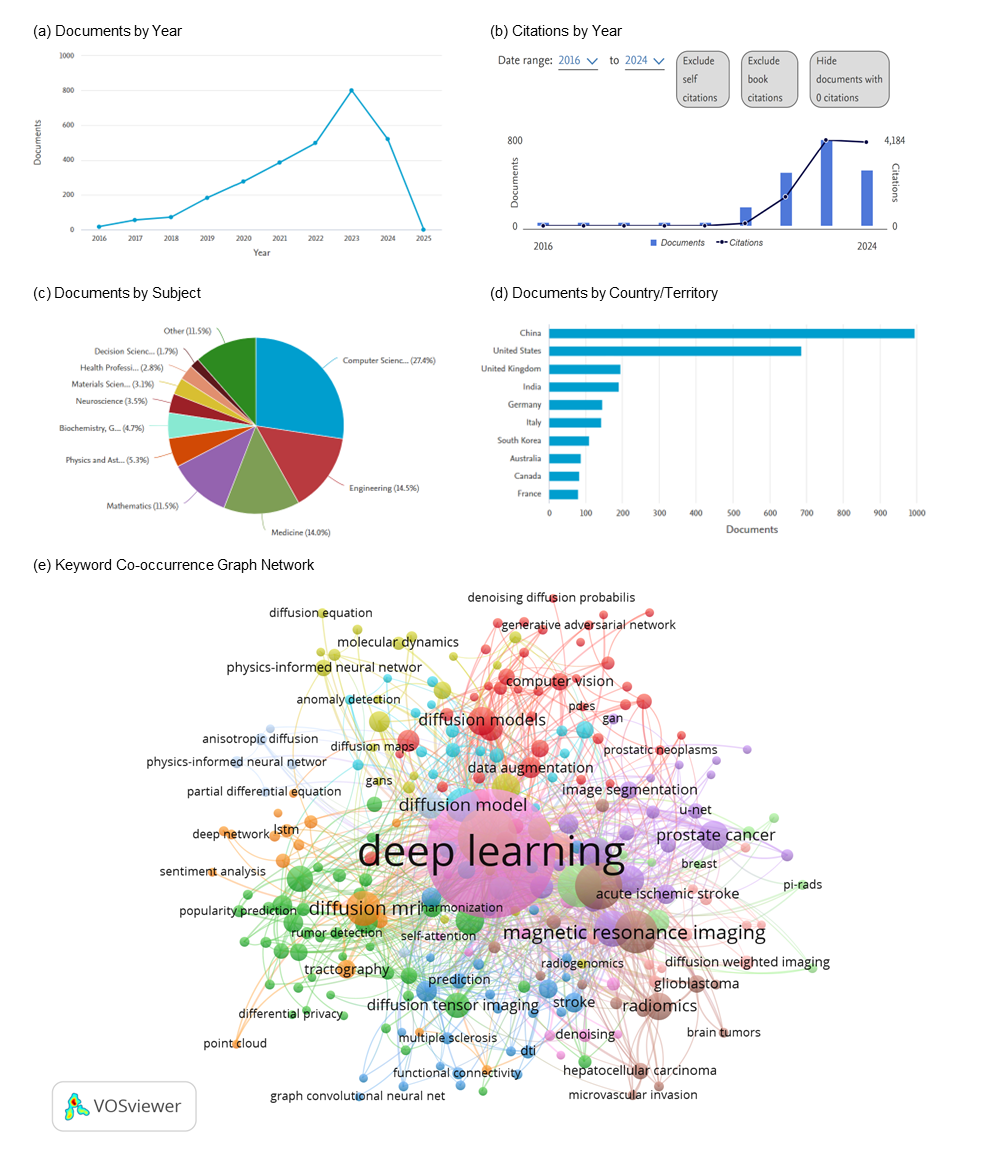}
    \caption{Scopus$^{\text{TM}}$ search results on diffusion-type generative AI models.
    (a) The number of diffusion-type generative AI papers;
    (b) the number of citations of these papers;
    (c) applications of diffusion-type generative AI models in various fields; 
    (d) contributions from top countries and territories; and
    (e) a VOSviewer$^{\text{TM}}$ presentation in terms of authors' key-words co-occurrences. With a threshold of 5 co-occurrences, 258 key words were selected to form this graph network.}
    \label{fig1}
\end{figure}

In preparing this review, we made our best effort in the literature search using various platforms and databases. In addition to Scopus and the GitHub repository we have mentioned above, we have also used PubMed, Google Scholar, arXiv, as well as proceedings of relevant conferences such as ICLR and NeurIPS. To visualize this booming field, in \Cref{fig1} (a)-(d) we show our Scopus$^{\text{TM}}$ search results with the rule in the field of Article Title, Abstract, and Key Words: “(Diffusion OR (Poisson AND flow)) AND (AI OR (Artificial AND Intelligence) OR (Deep AND Learning)) AND (Deep AND Network)”, which yielded 2,811 documents in total on June 28, 2024. Then, using the open-source software tool VOSviewer$^{\text{TM}}$, we further analyze the bibliometric results in terms of co-occurrence of key words to produce \Cref{fig1} (e).

The rest of this review is organized as follows. In the second section, physics-driven GMs are critically described from DMs to PFGM++, along with cutting-edge techniques for accelerated inference. In the third section, medical imaging applications of these GMs are concisely summarized, involving image reconstruction, image generation, and image analysis. In the last section, future directions are openly discussed, as related to unification of physics-inspired GMs, integration with VLMs, and potential novel applications of GMs.
    
    % physics-inspired generative models
    
\section{PHYSICS-INSPIRED GENERATIVE MODELS}
\label{sec:phys-gms}
%\subsection{Overview}
In this section, we give an overview of physics-inspired GMs, going from the earlier versions of DMs based on Markov chains all the way to the latest developments employing a combination of a score matching loss, distillation loss, and an adversarial loss to achieve state-of-the-art image quality and sampling efficiency. In doing so, we will place an particular emphasis on the generalized framework referred to as PFGM++~\cite{xu2023}, as this takes a key step towards the unification of physics-inspired GMs. 

\subsection{DDPMs}
\label{subsec:ddpm}
DDPMs~ \cite{sohl-dickstein2015,ho2020} are a class of GMs that have gained significant attention for their ability to generate high-quality images without adversarial training. The core idea behind DDPMs is to model the generation process as an iterative denoising process over a sequence of time steps, starting from pure noise and progressively adding structure to reach a sample from the data distribution. More formally, for each training data point $\bm{x}_0 \sim p(\bm{x})$ a discrete Markov chain ${\bm{x}_0,\bm{x}_1,...,\bm{x}_N}$ is constructed with transition kernel $p(\bm{x}_i|\bm{x}_{i-1})=\mathcal{N}(\bm{x}_i;\sqrt{\alpha_i}\bm{x}_i,(1-\alpha_i)\bm{I})$ for $\alpha_i := \prod_{j=1}^i(1-\beta_j).$ A sequence of positive noise scales $0<\beta_1,\beta_2,...,\beta_N<1$ will be chosen such that $\bm{x}_N$ are approximately distributed as $\mathcal{N}(\bm{0},\bm{I})$. The reverse variational Markov chain is parameterized as $p_{\bm{\theta}}(\bm{x}_{i-1}|\bm{x}_i)=\mathcal{N}(\bm{x}_{i-1}; \frac{1}{\sqrt{1-\beta_i}}(\bm{x}_i+\beta_is_{\bm{\theta}}(\bm{x}_i,i),\beta_i \bm{I}))$ and is learned through the objective: 
\begin{equation}
    \bm{\theta}^* = \argmin_{{\bm{\theta}}} \sum_{i=1}^N (1-\alpha_i) \mathbb{E}_{p(\bm{x})} \mathbb{E}_{p_{\alpha_i}(\tilde{\bm{x}}|\bm{x})} [||s_{\bm{\theta}}(\tilde{\bm{x}},i)-\nabla_{\tilde{\bm{x}}}\log p_{\alpha_i}(\tilde{\bm{x}} | \bm{x})||_2^2],
    \label{ddpm_obj}
\end{equation}
where $p_{\alpha_i}(\tilde{\bm{x}}):=\int p(\bm{x})p_{\alpha_i}(\tilde{\bm{x}}|\bm{x})d\bm{x}$ denotes the perturbed data distribution. Once equipped with a estimation of the minimizer in \Cref{ddpm_obj} a sample is generated by the reverse Markov chain 
\begin{equation}
    \bm{x}_{i-1} = \frac{1}{\sqrt{1-\beta_i}}(\bm{x}_i+\beta_is_{\bm{\theta}^*}(\bm{x}_i,i))+\sqrt{\beta_i}\bm{z}_i, i=N,N-1,...,1,
\end{equation}
starting from $\bm{x}_N \sim \mathcal{N}(\bm{0},\bm{I})$. The key to the success of DDPMs is that they transform the problem of generating samples from complex data distributions into a sequence of simpler denoising tasks, enabling high-quality image generation without adversarial training. 

\subsection{NCSNs}
\label{subsec:ncsn}
NCSNs~\cite{song2019} are a type of GM that leverage score-based generative modeling to produce high-quality synthetic samples. The fundamental concept of NCSNs involves learning the score of the data distribution conditioned on different noise scales, rather than directly learning the distribution itself. As for DDPM, let $p_\sigma(\bm{x}):=\int p(\bm{x})p_{\sigma}(\tilde{\bm{x}}|\bm{x})d\bm{x}$ denote the perturbed data distribution and $p_\sigma(\tilde{\bm{x}}|\bm{x}):=\mathcal{N}(\bm{x},\sigma^2\bm{I})$ the perturbation kernel. We then choose a sequence of positive noise scales $\sigma_{\text{min}}=\sigma_1, \sigma_2, ... \sigma_N = \sigma_{\text{max}}$ such that $p_{\sigma_{\text{min}}}(\bm{x})\approx p(\bm{x})$ and $p_{\sigma_{\text{max}}}(\bm{x})\approx \mathcal{N}(\bm{x};\bm{0},\sigma_{\text{max}}^2 \bm{I})$. NCSNs then estimate a score network $s_{\bm{\theta}}(\bm{x},\sigma)$ via the denoising score matching (DSM) \cite{vincent2011} objective
\begin{equation}
    \bm{\theta}^* = \argmin_{{\bm{\theta}}} \sum_{i=1}^N \sigma^2_i \mathbb{E}_{p(\bm{x})} \mathbb{E}_{p_{\sigma_i}(\tilde{\bm{x}}|\bm{x})} [||s_{\bm{\theta}}(\tilde{\bm{x}},\sigma_i)-\nabla_{\tilde{\bm{x}}}\log p_{\sigma_i}(\tilde{\bm{x}} | \bm{x})||_2^2]. 
\end{equation}
Using the estimated score network, a sample is generated by the $M$-step Langevin MCMC sampler 
\begin{equation}
    \bm{x}_i^m = \bm{x}_i^{m-1}+\epsilon_i s_{\bm{\theta}^*}(\bm{x}_i^{m-1},\sigma_i)+\sqrt{2\epsilon_i}\bm{z}_i^m, m=1,2,...,M,
\end{equation}
where $\epsilon_i>0$ denotes the step size, starting from an initial sample $\bm{x}_N \sim \mathcal{N}(\bm{0},\sigma_{\text{max}}^2 \bm{I})$.

\subsection{SDMs}
\label{subsec:SDM}
\begin{figure}
    \centering
    \includegraphics[width=\linewidth]{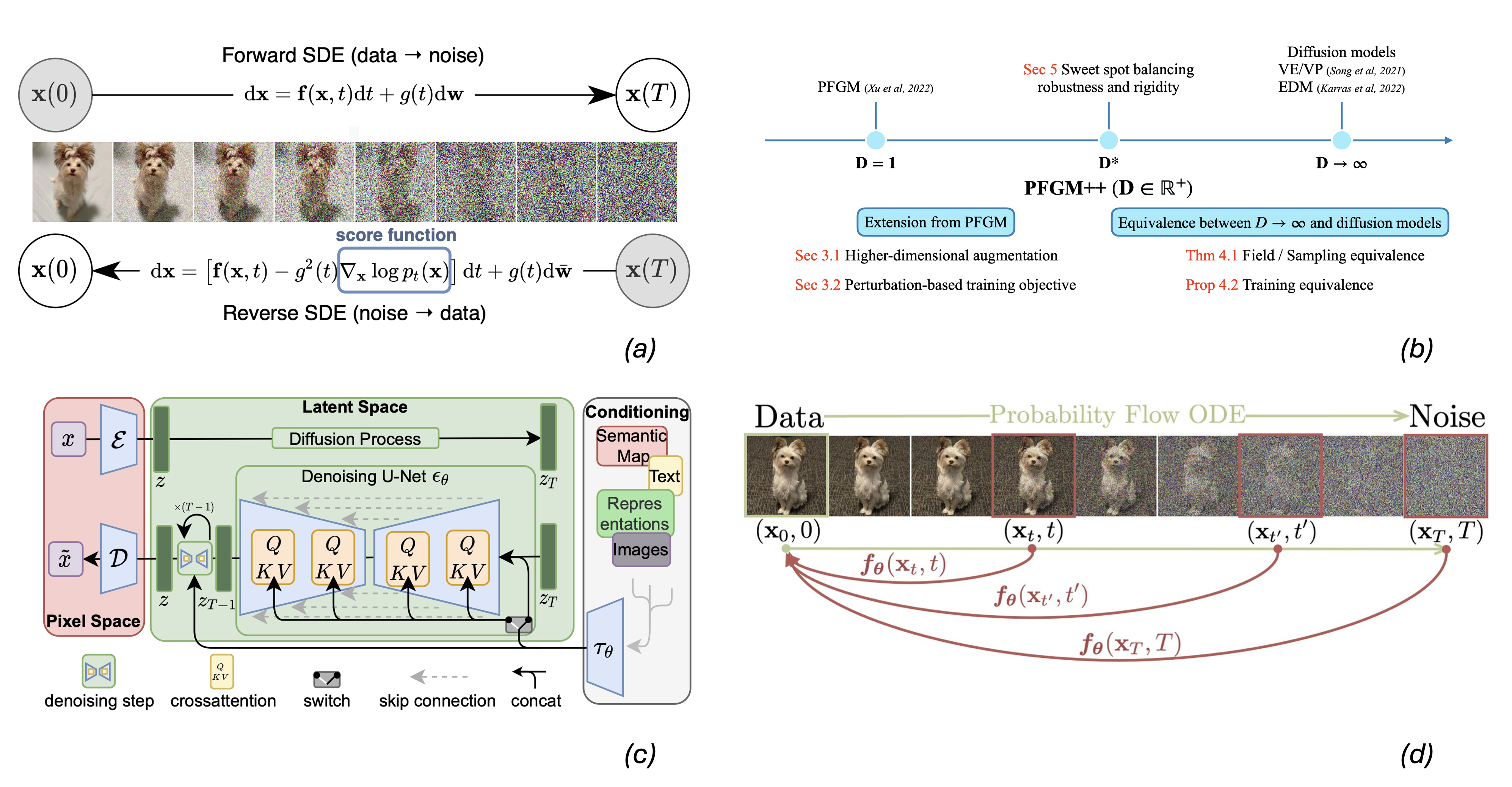}
    \caption{Physics-inspired deep GMs. 
    (a) Schematic of forward and reverse SDE on SDMs \cite{song2021};
    (b) Overview of PFGM++ \cite{xu2023}; 
    (c) Schematic overview of LDM \cite{rombach2022}; 
    (d) Schematic of mapping learned by consistency models \cite{song2023}. Note that different notation may have been used in the text compared to this figure.}
    \label{diff_fig}
\end{figure}

SDMs~\cite{song2021}, also known as score-based GMs, provide a unified framework for score-based models~\cite{song2019} and DMs~\cite{sohl-dickstein2015, ho2020} by formulating the diffusion process as a stochastic differential equation (SDE). Intuitively, this takes the idea of perturbing the data and learning to model complex data distributions via a sequence of simpler denoising task to the limit where the number of noise scales becomes infinite, leveraging advances in score modelling to obtain an effective sequential denoiser. This SDE takes can be written on the general form 
\begin{equation}
    d\bm{x} = \bm{\mu}(\bm{x},t)dt+g(t)d\bm{w},
\end{equation}
where $\bm{\mu}(\cdot, t)$ and $g(\cdot)$ are the drift and diffusion coefficients, respectively, and $\bm{w}$ is the standard Wiener process. One can choose the drift and diffusion coefficients, and $T$, such that $\bm{x}_T$ is some unstructured, easy to sample, noise distribution. Amazingly, according to the Anderson's theorem \cite{anderson1982}, the forward diffusion process has a corresponding reverse-time diffusion process, also a SDE, on the form
\begin{equation}
    d\bm{x} = [\bm{\mu}(\bm{x},t)-g(t)^2\nabla_{\bm{x}}\log p_t(\bm{x})]dt+g(t)d\bar{\bm{w}},
\end{equation}
where $\bar{\bm{w}}$ now is the standard Weiner process in reverse time. Crucially, the reverse-time SDE only depends on the data distribution via the gradient of the log-likelihood, known as the score function. An illustration of the forward and reverse processes is available in \Cref{diff_fig} (a). Corresponding to the reverse-time SDE, there is an ODE called the probability flow (PF) ODE \cite{song2021}:
\begin{equation}
    d\bm{x} = \left[\bm{\mu}(\bm{x},t)-\frac{1}{2}g(t)^2\nabla_{\bm{x}}\log p_t(\bm{x})\right]dt.
    \label{song_ode}
\end{equation}
As with NSCNs, the score function estimated via a DSM objective. Notably, as stated above, this frameworks includes DDPM and NCSN as special cases as they can be regarded as discretizations of two particular SDEs.  

The framework in \citea{karras2022}, which further builds on \cite{song2021}, describes the sampling process as moving along the PF ODE
\begin{equation}
    d\bm{x} = - \dot{\sigma}(t)\sigma(t)\nabla_{\bm{x}}\log p_{\sigma(t)} (\bm{x}) dt, 
    \label{edm_ode}
\end{equation}
where $\sigma(t)$ is a predefined, time-dependent, noise scale and $\nabla_{\bm{x}}\log p_{\sigma(t)} (\bm{x})$ is the time-dependent score function of the perturbed data distribution. This PF ODE corresponds to specific choices in the more general from presented in \Cref{song_ode}. In particular, we can move from \Cref{song_ode} to \Cref{edm_ode} by setting $\bm{\mu}(\cdot,t)=\bm{0}$, $\sigma(t)=t$, and $g(t)=\sqrt{2t}.$ Moving the ODE forward and backward in time, respectively, nudges the sample away from and towards the data distribution. Let $p(\bm{y})$ denote the data distribution, $p(\sigma)$ the training distribution of noise scales, $\lambda(\sigma)$ a weighting function, and $p_{\sigma}(\bm{x}|\bm{y})=\mathcal{N}(\bm{x},\sigma^2\bm{I})$ a Gaussian perturbation kernel, which samples perturbed versions $\bm{x}$ of ground truth data $\bm{y}.$ Then, the time-dependent score function is learned via the DSM objective, or perturbation based, objective
\begin{equation}
    \mathbb{E}_{\sigma \sim p(\sigma)} \mathbb{E}_{\bm{y}\sim p(\bm{y
    })} \mathbb{E}_{\bm{x}\sim p_{\sigma}(\bm{x}|\bm{y})} \left[\lambda(\sigma)||f_{\theta}(\bm{x},\sigma)-\nabla_{\bm{x}} \log p_{\sigma}(\bm{x}|\bm{y})||_2^2\right]. 
    \label{edm_obj}
\end{equation}
Once equipped with this estimate, we can generate an image by drawing an initial sample from the prior noise distribution and solving \Cref{edm_ode} using some numeric ODE solver. SDMs offer a powerful and elegant framework for generative modeling, leveraging the mathematics of diffusion processes and score functions to transform noise into high fidelity samples. 

\subsection{PFGMs}
\label{subsec:pfgm}
PFGMs~\cite{xu2022} are a relatively new family of deep GMs inspired by electrostatics, instead of non-equilibrium thermodynamics as is the case for DMs. In particular, PFGMs treat the $N$-dimensional data $\bm{x} \in \mathbb{R}^N$, as electric charges in a $N+1$-dimensional augmented space $\bm{\tilde{x}}:=(\bm{x},z) \in \mathbb{R}^{N+1}$. Following the law of motion defined by the electric field line emitted by the charges, the data distribution placed on the $z=0$ hyperplane can be transformed into a distribution on the hemisphere with radius $r.$ As $r \rightarrow \infty$ this distribution will become uniform \cite{xu2022}. Instead of estimating a time-dependent score function, as in SDMs, PFGMs learn an empirical Poisson field, derived from the Poisson equation in the augmented space,
\begin{equation}
    \bm{E} (\bm{\tilde{x}}) = \frac{1}{S_N(1)} \int \frac{\bm{\tilde{x}}-\bm{\tilde{y}}}{||\bm{\tilde{x}}-\bm{\tilde{y}}||^{N+1}} p(\bm{y}) d\bm{y},
    \label{pfgm_E}
\end{equation}
where $p(\bm{y})$ is the data distribution and $S_N(1)$ denotes the surface area of a unit N-sphere. We can move from the data distribution to the uniform distribution on the infinite hemisphere by running the ODE $d\bm{\tilde{x}} = \bm{E} (\bm{\tilde{x}}) dt$ forward and back in time. In practice, PFGMs are trained using a large batch of size $n$ to approximate the integral in \Cref{pfgm_E} to estimate a normalized empirical Poisson field $\bm{\hat{E}}(\bm{\tilde{x}}) = c(\bm{\tilde{x}}) \sum_{i=1}^n \frac{\bm{\tilde{x}}-\bm{\tilde{y}}_i}{||\bm{\tilde{x}}-\bm{\tilde{y}}_i||^{N+1}}.$ The training objective is to minimize the $\ell_2$-loss between the network $f_\theta(\bm{\tilde{x}})$ and $\bm{E}(\bm{\tilde{x}})/||\bm{E}(\bm{\tilde{x}})||$ over various, carefully selected to ensure coverage of sampling trajectories, positions of $\bm{\tilde{x}}.$ PFGMs represent a powerful approach to generative modeling, combining the stochastic modeling capabilities of Poisson processes with flow-based generative techniques. 

\subsection{PFGM++}
\label{subsec:pfgm++}
PFGM++~\cite{xu2023} take an important step towards unification of physics-inspired GMs since it contains PFGMs \cite{xu2022} and, remarkably, SDMs~\cite{song2021,karras2022} as special cases. An overview is available in \Cref{diff_fig} (b). PFGM++ in essence generalize PFGMs in terms of augmented space dimensionality. The $N$-dimensional data, $\bm{x}\in\mathbb{R}^N$, are treated as positive electric charges in a $N+D$-dimensional augmented space. Let $\tilde{\bm{x}}:=(\bm{x},\bm{z}) \in \mathbb{R}^{N+D}$ and $\tilde{\bm{y}}:=(\bm{x},\bm{0}) \in \mathbb{R}^{N+D}$ denote the augmented perturbed and ground truth data, respectively. The electric field lines generated by these charges define a surjection via the ODE, $d\bm{\tilde{x}} = \bm{E} (\bm{\tilde{x}}) dt$, between an uniform distribution on the infinite $N+D$-dimensional hemisphere and the data placed on the $\bm{z}=\bm{0}$ hyperplane. The objective of interest in PFGM++ is the high dimensional electric field
\begin{equation}
    \bm{E} (\bm{\tilde{x}}) = \frac{1}{S_{N+D-1}(1)} \int \frac{\bm{\tilde{x}}-\bm{\tilde{y}}}{||\bm{\tilde{x}}-\bm{\tilde{y}}||^{N+D}} p(\bm{y}) d\bm{y},
    \label{pfgmpp_E}
\end{equation}
where $S_{N+D-1}(1)$ is the surface area of the unit $(N+D-1)$-sphere, and $p(\bm{y})$ the ground truth data distribution. \citea{xu2023} note that the electric field is rotationally symmetric on the $D$-dimensional cylinder $\sum_{i=1}^D z_i^2 = r^2, \forall r > 0$ and it suffices to track the norm of the augmented variables $r(\bm{\tilde{x}}):=||\bm{z}||_2.$ In particular, we can observe that
\begin{equation}
    \frac{dr}{dt} = \sum_{i=1}^D \frac{z_i}{r}\frac{dz_i}{dt} = \frac{1}{S_{N+D-1}(1)} \int \frac{r}{||\bm{\tilde{x}}-\bm{\tilde{y}}||^{N+D}} p(\bm{y}) d \bm{y}, 
\end{equation}
where  $d z_i = \bm{E} (\bm{\tilde{x}})_{z_i}dt$, and $\bm{E} (\bm{\tilde{x}})_{z_i}$ denotes the $z_i$ component of $\bm{E} (\bm{\tilde{x}}).$ 
 For notational brevity, we can redefine $\tilde{\bm{y}}:=(\bm{y},0) \in \mathbb{R}^{N+1}$ and $\tilde{\bm{x}}:=(\bm{x},r) \in \mathbb{R}^{N+1}$. The ODE of interest is then 
\begin{equation}
    d\bm{x} = \bm{E} (\bm{\tilde{x}})_{\bm{x}} \cdot E(\bm{\tilde{x}})_r^{-1} \label{pfgmpp_ode} dr, 
\end{equation}
where $E(\bm{\tilde{x}})_r=\frac{1}{S_{N+D-1}(1)} \int \frac{r}{||\bm{\tilde{x}}-\bm{\tilde{y}}||^{N+D}}p(\bm{y})d\bm{y}$, a scalar. Crucially, this symmetry reduction has converted the aforementioned surjection into a bijection between the ground truth data placed on the $r=0$ $(\bm{z}=\bm{0})$ hyperplane and a distribution on the $r=r_{\max}$ hyper-cylinder \cite{xu2023}. For formal statement and proof see Theorem 3.1. in \citea{xu2023}. PFGM++ address several drawbacks of the training objective used in PFGMs by proposing a perturbation based objective, inspired by the DSM objective successfully used in SDMs \cite{song2021,karras2022}. Let $p_r(\bm{x} | \bm{y})$ denote the perturbation kernel and $p(r)$ the training distribution over $r$. Then the objective can be formulated as
\begin{equation}
    \mathbb{E}_{r\sim p(r)} \mathbb{E}_{p(\bm{y})} \mathbb{E}_{p_r(\bm{x}|\bm{y})} \left[ ||f_\theta (\bm{\tilde{x}})-\frac{\bm{x}-\bm{y}}{r/\sqrt{D}} ||_2^2\right]. 
    \label{pfgmpp_obj_final}
\end{equation}
For $p_r(\bm{x}|\bm{y}) \propto 1/(|| \bm{x}-\bm{y}||_2^2+r^2)^{\frac{N+D}{2}}$, it is possible to show that the minimizer of \Cref{pfgmpp_obj_final} is $f^*_\theta(\bm{\tilde{x}}) = \sqrt{D} \bm{E} (\bm{\tilde{x}})_{\bm{x}} \cdot E(\bm{\tilde{x}})_r^{-1}.$ Since the perturbation kernel is isotropic, one can sample from $p_r(\cdot|\bm{y})$ by splitting it into hyperspherical coordinates to $\mathcal{U}_\psi(\psi)p_r(R)$, where $\mathcal{U}_\psi$ denotes the uniform distribution of the angle component and the distribution of $R=||\bm{x}-\bm{y}||$, the radius, is 
\begin{equation}
    p_r(R)\propto \frac{R^{N-1}}{(R^2+r^2)^{\frac{N+D}{2}}}.
\end{equation}
Once equipped with the minimizer of \Cref{pfgmpp_obj_final}, we can generate a sample from our target distribution by solving $d\bm{x}/dr = \bm{E}(\bm{\tilde{x}})_{\bm{x}}/E(\bm{\tilde{x}})_r = f^*_\theta(\bm{\tilde{x}})/\sqrt{D}$ with an initial sample from $p_{r_{\max}}.$ 

\subsubsection{PFGM++ and SDMs in theory and in practice}
PFGM++ admit PFGMs~\cite{xu2022} ($D=1$) and, remarkably, SDMs~\cite{song2021,karras2022} ($D \to \infty$) as special cases. In addition to being intimately connected in theory, PFGM++ and SDMs \cite{song2021,karras2022} are also closely connected in practice. \citea{xu2023} demonstrate that one can re-purpose the training and sampling algorithms developed for DMs in \cite{karras2022} to PFGM++. This follows from the fact that $\sigma:=\sigma(t)=t$ in \cite{karras2022}, $\bm{\tilde{x}}:=(\bm{x},r)$, and the alignment formula $r=\sigma \sqrt{D}$, which allows for the change-of-variables $d\bm{x} = f^*_\theta(\bm{\tilde{x}})/\sqrt{D} dr = f^*_\theta(\bm{\tilde{x}}) dt$ since $dr=d\sigma \sqrt{D}=dt \sqrt{D}.$ 

\subsubsection{Trade-off robustness for rigidity by varying $D$}
In PFGM++, $D$ is treated as a hyperparameter to be tuned. In particular,~\citea{xu2023} demonstrate that by tuning $D\in(0,\infty)$ one can trade off robustness (insensitive to missteps) for rigidity (effective learning). In particular, as first shown for PFGM \cite{xu2022}, PFGM++ are less susceptible to estimation errors than SDMs.~\citea{xu2023} demonstrate the superior robustness of PFGM++ to three sources of errors: noise injection, large step-size, and post-training quantization.  At training time, for large $D$ and $r$, the marginal distribution $p_r(\bm{x})$ is supported on a sphere with radius $r\sqrt{N/D}.$ If the sample trajectory deviates from this $r$-norm relation in the training data, the backward ODE can lead to unexpected results. Hence, setting $D$ small, moving away from SDMs, increase robustness. However, a small $D$ results in a heavy-tailed input distribution and a broad input range is challenging to handle for any finite capacity network. Thus, optimal performance occurs for some $D\in (0,\infty)$ that requires tuning for the problem at hand.

\subsection{Latent DMs (LDMs)}
\label{subsec:ldm}
LDMs~\cite{rombach2022} innovate by operating in a latent space—a compressed representation of the data—rather than directly in the data's original space. This method significantly reduces computational costs while maintaining, or even enhancing, the quality of generated samples. In particular,~\cite{rombach2022} suggest a perceptual image compression via an autoencoder trained with a combination of perceptual and adversarial objectives. For a RGB image $\bm{x}\in \mathbb{R}^{H\times W\times 3}$ the encoder $\mathcal{E}$ encodes $\bm{x}$ into the latent representation $\bm{z}=\mathcal{E}(\bm{x}) \in \mathbb{R}^{h\times w\times c}.$ The decoder $\mathcal{D}$ reverse this process, reconstructing the image from its latent representation $\tilde{\bm{x}}=\mathcal{D}(\mathcal{E}(\bm{x})).$ \citea{rombach2022} investigate a range of down-sampling factors $f=H/h=W/h$ and finds that $f=\{4,8\}$ yields a good trade-off between computational complexity and image fidelity. In addition,~\citea{rombach2022} introduce a very general conditioning mechanism to model conditional distributions $p(\bm{z}|\bm{y})$ by augmenting the DM backbone with a cross-attention mechanism \cite{vaswani2017}, allowing the network to efficiently learn embeddings from varying modalities via domain specific encoders $\tau_{\bm{\theta}}(y)\in \mathbb{R}^{M\times d_\tau}.$ A schematic overview of LDMs is available in \Cref{diff_fig} (c). 

LDMs are celebrated for their ability to produce high-quality outputs with reduced computational demands. By leveraging the efficiencies of latent space, LDMs offer a powerful solution for generating complex data, such as high-resolution images or intricate text sequences, making them a valuable tool in the arsenal of generative modeling techniques. A particular instance of LDMs, a text-to-image DM called Stable Diffusion~( \url{https://github.com/CompVis/stable-diffusion}) has gained unprecedented popularity for its effectiveness in image generation and manipulation.

\subsection{Accelerated Sampling Schemes}
\label{subsec:ddim}
The iterative nature of physics-inspired GMs is a double-edge sword. On the one hand, it allows for a flexible trade-off between compute and image quality, in addition to zero-shot inverse problem solving. On the other hand, this means that the number of function evaluations (NFE) required may be order of magnitudes larger than for single-step samplers. This computationally heavy sampling limit their usage in applications where speed is of the essence. 

One line of work to accelerate sampling involves developing more efficient samplers. In particular, a salient trend is to move from stochastic to deterministic samplers. This was done for DDPMs in Denoising Diffusion Implicit Models (DDIMs)~\cite{song2020} and for SDMs in~\citea{karras2022}. Notably, relatively newer methods such as PFGMs \cite{xu2022} and PFGM++~\cite{xu2023} also follow this trend, focusing directly on the ODE instead of the SDE. 

Another prominent development in the quest for more efficient sampling are consistency models \cite{song2023,song2023a}. Consistency models are a class of GMs that build on the continuous-time formulation of DMs \cite{song2021,karras2022} to enable single-step, $\text{NFE}=1$, sampling. These models achieve this by directly mapping noise to data, enabling high-quality sample generation in a single step, by enforcing self consistency at training time. An illustration of the mapping learned in consistency models is available in \Cref{diff_fig} (d). Let $\bm{x}_\sigma$ denote an intermediate step on the solution trajectory $\{\bm{x}_\sigma\}_{\sigma\in[\sigma_{\text{min}},\sigma_{\text{max}}]}$, where $\bm{x}_{\sigma_{\text{min}}}\sim p(\bm{x})$ and $\bm{x}_{\sigma_\text{max}} \sim \mathcal{N}(\bm{0},\sigma^2_{\text{max}}).$ Consistency models then strive to learn a consistency function
\begin{equation}
    f (\bm{x}_\sigma,\sigma) \mapsto \bm{x}_{\sigma_{\text{min}}}
    \label{consistency_function}
\end{equation}
by enforcing self-consistency
\begin{equation}
    f(\bm{x}_\sigma,\sigma)=f(\bm{x}_{\sigma'},\sigma'), \forall \sigma,\sigma' \in [\sigma_{\text{min}},\sigma_{\text{max}}]. 
\end{equation}
One can ensure that the desired solution is the consistency function in \Cref{consistency_function} by enforcing the boundary condition
\begin{equation}
    f_{\bm{\theta}}(\bm{x},\sigma)=c_{\text{skip}}(\sigma)\bm{x}+c_{\text{out}}(\sigma) F_{\bm{\theta}}(\bm{x},\sigma), 
\end{equation}
where $c_{\text{skip}}(\sigma)$ and $c_{\text{out}}(\sigma)$ are differentiable functions such that $c_{\text{skip}}(\epsilon)=1$ and $c_{\text{out}}(\epsilon)=0$ and $f_{\bm{\theta}}(\bm{x},\sigma)$ denotes our approximated consistency function via the deep neural network $F_{\bm{\theta}}(\bm{x},t).$ The domain $[\sigma_{\text{min}},\sigma_{\text{max}}]$ is discretized using a sequence of noise scales $\sigma_{\text{min}}=\sigma_1 < \sigma_2 < ... < \sigma_N = \sigma_{\text{max}}.$ Following~\citea{karras2022},~\citea{song2023} set $\sigma_i=(\sigma_{\text{min}}^{1/\rho}+\frac{i-1}{N-1}(\sigma_{\text{max}}^{1/\rho}-\sigma_{\text{min}}^{1/\rho}))^\rho$ with $\rho=7$ for $i=1,...,N.$ 

Consistency models are trained via the consistency matching loss
\begin{equation}
    \mathbb{E} \left[\lambda(\sigma_i) d(f_{\bm{\theta}}(\bm{x}_{\sigma_{i+1}},\sigma_{i+1}),f_{\bm{\theta}^-}(\breve{\bm{x}}_{\sigma_i},\sigma_{i}))\right], 
    \label{cm_loss}
\end{equation}
where 
\begin{equation}
    \breve{\bm{x}}_{\sigma_{i}}=\bm{x}_{\sigma_{i+1}}-(\sigma_i-\sigma_{i+1})\sigma_{i+1}\nabla_{\bm{x}} \log p_{\sigma_{i+1}}(\bm{x})|_{\bm{x}=\bm{x}_{\sigma_{i+1}}}, 
    \label{ode_solve}
\end{equation}
denotes a single-step in the reverse direction of the PF ODE in \Cref{edm_ode}, $d(\cdot,\cdot)$ a metric function, and $\lambda(\cdot)\in \mathbb{R}_+$  a positive weighting function. The expectation in \Cref{cm_loss} is taken with respect to $i\sim \mathcal{U}[1,N-1]$, a uniform distribution over integers $i=1,...,N-1$, and $\bm{x}_{\sigma_{i+1}} \sim p_{\sigma_{i+1}}(\bm{x}).$ The objective in \Cref{cm_loss} is minimized via stochastic gradient descent on parameters $\bm{\theta}$ whilst $\bm{\theta}^-$ are updated via the exponential moving average (EMA)
\begin{equation}
    \bm{\theta^-} \leftarrow \text{stopgrad}(\mu\bm{\theta^-}+(1-\mu)\bm{\theta}), \;
\end{equation}
where $0\leq \mu \leq 1$ is the decay rate. Since $\nabla_{\bm{x}}\log p_{\sigma_{i+1}}(\bm{x})$ is not known, minimizing the objective in \Cref{cm_loss} is not feasible. Consistency training and consistency distillation are two algorithms to address this issue \cite{song2023}. Both approaches use an approximation to generate a pair of adjacent points on the solution trajectory, approximately adhering to \Cref{ode_solve}. Consistency distillation, as the name implies, distills a pre-trained DM into a consistency model and simply replacing the score-function in \Cref{ode_solve} by a pre-trained score network. Consistency training, on the other hand, is a stand-alone model and estimates the two adjacent points on the solution trajectory via the perturbation kernel. In particular, consistency training obtains the pair of adjacent points on the same solution trajectory $(\bm{x}_{\sigma_{i+1}},\breve{\bm{x}}_{\sigma_i})$ via $\bm{x}_{\sigma_{i+1}}=\bm{x}+\sigma_{i+1}\bm{z}$ and $\breve{\bm{x}}_{\sigma_i}=\bm{x}+\sigma_i \bm{z}$ using the same $\bm{x}$ and $\bm{z}.$ Consistency training then employ the loss
\begin{equation}
    \mathbb{E} \left[\lambda(\sigma_i) d(f_{\bm{\theta}}(\bm{x}+\sigma_{i+1}\bm{z},\sigma_{i+1}),f_{\bm{\theta}^-}(\bm{x}+\sigma_{i}\bm{z},\sigma_{i}))\right], 
    \label{cd_loss}
\end{equation}
which is equivalent to the consistency matching loss in \Cref{cm_loss} in the $N\rightarrow \infty$ limit \cite{song2023}. After approximating the consistency function, generating a sample involves sampling an initial noise vector from the noise distribution and applying the consistency model: $\hat{\bm{x}}=f_{\bm{\theta}}(\bm{z},\sigma_{\text{max}})$. Importantly, consistency models maintain the flexibility of balancing image quality against computational efficiency and enable zero-shot data editing, supporting the option of multi-step sampling processes for enhanced control over output quality \cite{song2023}. 

Using self-consistency to achieve high fidelity output with significantly less compute has been extended to the case of LDM in \citea{luo2023}, with a method called Latent Consistency Distillation (LCD). In addition to generalizing the consistency matching framework to text-to-image LDM, the authors also generalizes the idea of enforcing consistency between adjacent time steps $t_{n+1}$ and $t_n$ to consistency between time steps that are $k$ steps apart, that is $t_{n+k}$ and $t_n.$ This methodology generates impressive state of the art few-step, $\text{NFE}<8$, text-to-image generation. 

Consistency trajectory models (CTMs)~\cite{kim2023} learn to estimate both the integrand (the score function) and the integral over any time horizon of the PF ODE, providing a unifying framework which includes consistency models and DMs as special cases. This setup enables anytime-to-anytime jumps, which is leveraged efficiently to trade off compute for image quality via novel sample techniques. In particular, CTMs achieve impressive state of the art performance for unconditional image generation on CIFAR-10 and ImageNet at $64\times64$ resolution datasets with $\text{NFE}=1$. Notably, however, CTMs are using a combination of a distillation loss, DSM loss, and an adversarial loss. The inclusion of an adversarial loss can lead to training instability, tentatively negating a key advantage of DMs over GANs.

    % applications in medical imaging
    % \newpage
\section{APPLICATIONS IN MEDICAL IMAGING}

The advent of physics-inspired GMs has offered new and exciting opportunities in medical imaging. By leveraging the physical principles, these models offer transformative potential in enhancing the performance of medical imaging tasks. In medical imaging, high-quality images are paramount for diagnosis, therapy, and follow-up. Deep GMs, described in \Cref{sec:phys-gms} excel in learning complex data priors, leveraging them to infer missing or corrupted information and generate high-fidelity images. This section delves into three primary applications: \Romannum{1}) Image Reconstruction, \Romannum{2}) Image Generation, and \Romannum{3}) Image Analysis.

\subsection{Image Reconstruction}

Major medical imaging modalities include CT, MRI, nuclear imaging (PET/SPECT), ultrasound, and optical imaging. The imaging challenges are modality-dependent, such as the radiation dose reduction in CT or the k-space measurement acceleration in MRI, both of which need to handle partial and/or noisy data for tomographic reconstruction.

The raw data acquisition in medical imaging can be modeled by an operator \(\mathcal{F} \in \mathbb{R}^{n} \mapsto \mathbb{R}^{m}\) acting on the continuous image domain \(\mathbf{x} \in \mathbb{R}^{n}\), resulting in a dataset in the measurement domain \(\mathbf{y} \in \mathbb{R} ^{m} = \mathcal{F}(\mathbf{x}) + \mathbf{n}\), where \(\mathbf{n} \in \mathbb{R} ^ {n}\) represents noise. 
In CT or PET, \(\mathcal{F}\)  corresponds to the Radon transform, which produces line integrals of an underlying image to be reconstructed. In SPECT, an attenuated Radon transform is involved.
In MRI, \(\mathcal{F}\) commonly corresponds to a multi-coil Fourier sampling operator.
In practice, the problem is commonly discretized as \(\mathbf{y} = \mA \mathbf{x} + \mathbf{n}\), where \(\mA \in \mathbb{R}^{m \times n}\). 

\begin{figure}
    \centering
    \includegraphics[width=\textwidth]{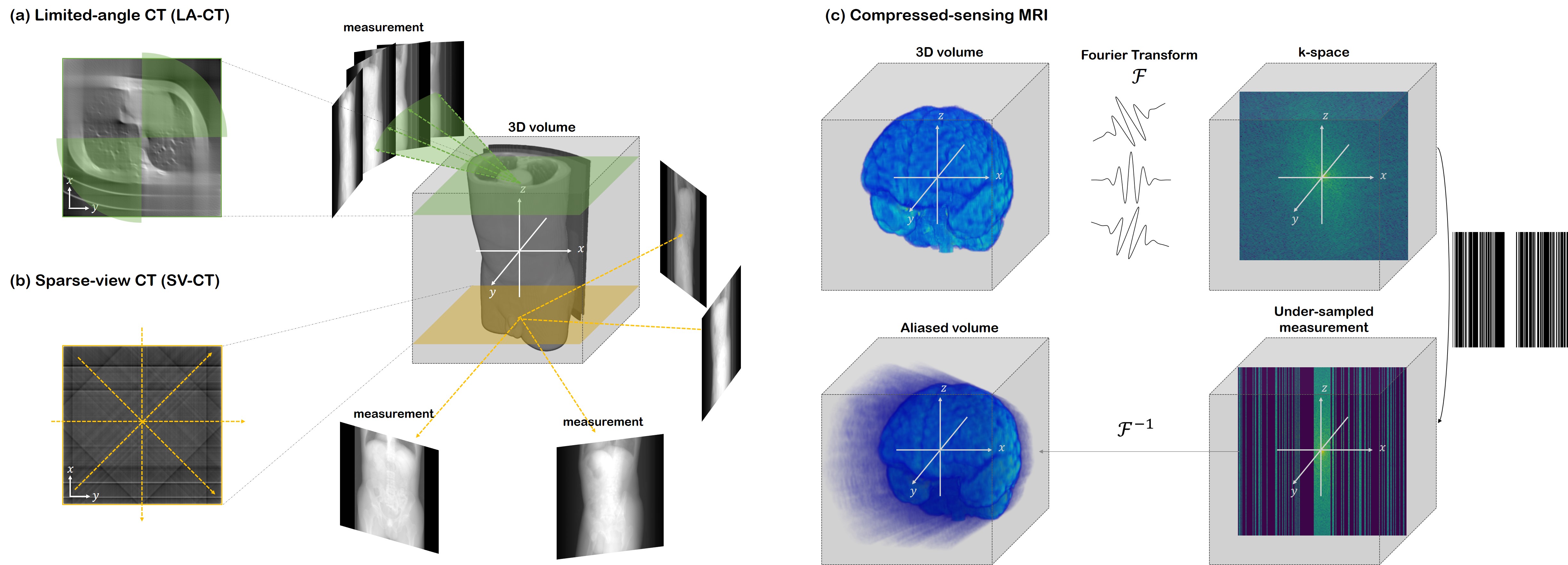}
    \caption{Visualization of three exemplary measurement processes for CT and MRI. (a) limited-angle CT (LA-CT); (b) sparse-view CT (SV-CT); and (c) compressed sensing-based MRI (CS-MRI). Adapted from \cite{chung2023solving}.}
    \label{fig:forward-model}
\end{figure}

The \(\mA\) can be ill-posed ($m < n$), especially when measurements are taken at sub-Nyquist rates, making direct recovery of \(\mathbf{x}\) rather challenging.
Taking CT and MRI as two examples, \Cref{fig:forward-model} visualizes the measurement techniques used in three specific imaging modes: limited-angle CT (LA-CT), sparse-view CT (SV-CT), and compressed sensing MRI (CS-MRI).

Compressed sensing theory~\cite{candes2006robust, donoho2006compressed} provides a robust foundation for improving sparse-view and low-dose CT reconstruction. Similarly, SENSE (Sensitivity Encoding)~\cite{pruessmann1999sense} is widely used for MRI reconstruction. In a nutshell, compressed sensing-based iterative reconstruction (IR) can be formulated as the optimization problem,
\begin{equation}
 \label{eq:ir}
 \hat{\mathbf{x}} = \arg \min_{\mathbf{x}} \| \mA \mathbf{x} - \mathbf{y} \|_2^2 + \lambda \mathcal{R}(\mathbf{x}),
\end{equation}
\noindent
where the first term ensures data consistency while \(\mathcal{R}(\mathbf{x})\) is a regularization term promoting sparsity, and \(\lambda\) is a regularization parameter. Techniques such as total variation (TV)~\cite{yu2005total}, anisotropic TV~\cite{chen2013limited}, total generalized variation (TGV)~\cite{niu2014sparse}, fractional order TV~\cite{zhang2016image} and many other variants, have all improved reconstruction results.

\Cref{eq:ir} is often minimized using iterative first-order optimization algorithms like gradient descent. Starting from \(\mathbf{x}_0\) with a step size \(\eta_t\), the \((t+1)^\text{th}\) iteration of gradient descent is written as,
\begin{equation}
    \mathbf{x}_{t+1} = \mathbf{x}_t - \eta_t \nabla \mathcal{L}(\mathbf{x}_t) = \mathbf{x}_t - \eta_t \left( \mA^T (\mA\mathbf{x}_t - \mathbf{y}) + \nabla \mathcal{R}(\mathbf{x}_t) \right),
\end{equation}
where \(\nabla \mathcal{R}(\mathbf{x}) = \nabla \log p(\mathbf{x})\) represents the score of the distribution, acting as a noise/artifact estimate. Subtracting this term helps refine the image, thus the gradient descent step effectively removes noise and/or aliasing components from the current estimate \(\mathbf{x}_t\).

Traditional image reconstruction methods have been extensively used, such as filtered backprojection (FBP)~\cite{fbp} and iterative reconstruction techniques like the algebraic reconstruction technique (ART)~\cite{gordon1970algebraic}, simultaneous algebraic reconstruction technique (SART)~\cite{andersen1984simultaneous}, and expectation maximization (EM)~\cite{dempster1977maximum}. However, these methods face limitations due to noise sensitivity, imperfect data challenges, and computational overhead. For example, in sparse-view CT the angular down-sampling fails to satisfy Nyquist's criterion~\cite{xia2022patch}, leading to unsatisfactory results with either analytical or compressed sensing-based iterative reconstruction algorithms, if the down-sampling is done too aggressively. 

Deep learning methods for CT image reconstruction have commonly employed supervised learning to map partial measurements to complete images~\cite{wei20202steps, mardani2017}, relying on extensive datasets of paired CT images and measurements. However, variations in the measurement process, such as changing CT projections, require new data collection and model retraining. This limits generalization and can sometimes lead to counter-intuitive instabilities~\cite{antun}. Recently, DMs have addressed these issues by introducing unsupervised training, which provides effective priors without the need for labeled data~\cite{song2021solving, chung2022diffusion}. These models, which train on high-quality images/data to fill in gaps in down-sampled data, eliminating the need for matched datasets, supporting unsupervised learning and improving tomographic reconstruction quality. The following sub-sections delve into the application of DMs in tomographic, MRI, and PET reconstruction, emphasizing their role in producing superior images and diagnostic performance.

\subsubsection{CT Reconstruction}

The initial work reported in \cite{song2021solving} adapted the diffusion model to medical image reconstruction. This score-based generative model, as outlined in \Cref{subsec:SDM}, involves solving inverse problems by generating samples from the posterior distribution \( p(\rvx \mid \rvy) \). This is achieved by conditioning a stochastic process \(\{\rvx_t\}_{t\in[0,1]}\) on an observation \(\rvy\), leading to the conditional process \(\{\rvx_t \mid \rvy\}_{t\in[0,1]}\). The core of this approach is the conditional stochastic process as a reverse-time SDE: 
\begin{equation}
    \ud \rvx_t = \left[ f(t)\rvx_t - g(t)^2 \nabla_{\rvx_t} \log p_t(\rvx_t \mid \rvy) \right] \ud t + g(t) \ud \bar{\rvw}_t,
\end{equation}
where the conditional score function \(\nabla_{\rvx_t} \log p_t(\rvx_t \mid \rvy)\) is approximated with an unconditionally trained score model \(\vs_{\vtheta^*}(\rvx, t)\) and the measurement distribution \( p(\rvy \mid \rvx) \). Notably, this is an unsupervised method and thus does not require paired data. 

The method utilizes a special formulation of the linear operator \(\mA\), decomposed as \(\mA = \mcal{P}(\bm{\Lambda}) \mT\), where \(\mT\) represents common transforms like Radon, and \(\bm{\Lambda}\) is a diagonal matrix indicating subsampling. This decomposition facilitates efficient sample generation from \(\{\rvx_t \mid \rvy\}_{t\in[0,1]}\) by leveraging an unconditional score model and incorporating the conditional information. The iterative sampling process modifies existing sampling algorithms by introducing a step that enforces consistency with the observation \(\rvy\). This is done by solving a proximal optimization problem:
\begin{equation}
    \hat{\rvx}_{t_i}' = \argmin_{\vz} \{ (1- \lambda) \|\vz - \hat{\rvx}_{t_i}\|_\mT^2 + \min_{\vu} \lambda \|\vz - \vu\|_\mT^2\}, \quad \text{s.t.} \quad \mA \vu = \hat{\rvy}_{t_i},
\end{equation}
where \(\lambda\) balances between fitting the sample to both the score model and the observation. The process allows the method to be widely applicable to various sampling methods for various medical image reconstruction tasks. In this study, the norm \(||\va_\mT^2|| = ||\mT \va_2^2||\) is chosen to simplify the analysis. The decomposition allows for a closed-form solution given by
\begin{equation}
    \hat{\rvx}_{t_i}' = \mT^{-1}[\lambda \bm{\Lambda} \mcal{P}^{-1}(\bm{\Lambda})\hat{\rvy}_{t_i} + (1 - \lambda) \bm{\Lambda} \mT \hat{\rvx}_{t_i} + (\mI - \bm{\Lambda}) \mT \hat{\rvx}_{t_i}]
\end{equation}

where $\mcal{P}^{-1}(\bm{\Lambda})$ is a right inverse that increases the dimensionality of a vector by padding entries according to a subsampling mask $\bm{\Lambda}$. When $\lambda=0$, the constraint $\mA \hat{\rvx}_{t_i}' = \hat{\rvy}_{t_i}$ is ignored, leading to unconditional generation. When $\lambda=1$, the constraint is exactly satisfied. For noisy measurements, $0 < \lambda < 1$ balances between fitting the data and the constraint. 

Along this direction, \citea{cong2023image} improved the diffusion model-based tomographic reconstruction approach to perform the MAP reconstruction in the Bayesian framework 
% \textcolor{red}{(cf. the introduction)} 
utilizing a Gaussian mixture model that more accurately captures discrepancies between the distributions of reconstructed images and real counterparts, respectively. This integration facilitates the derivation of a novel score-matching formula implemented via deep learning.
The performance of this Gaussian mixture score-matching method was evaluated on both public medical image datasets and clinical raw CT datasets, demonstrating its effectiveness and robustness compared with traditional methods~\cite{tang2009performance} and existing DMs.

While traditional approaches to transmission tomography often rely on linearizing the reconstruction problem, leading to inaccuracies with complex issues, \citea{li2024ct} introduced a nonlinear method called Diffusion Posterior Sampling (DPS). This method performs CT image reconstruction by integrating a score-based diffusion model with a nonlinear physics-driven measurement likelihood. The DPS  method transforms the reconstruction process by applying the Bayesian rule to merge a pre-trained diffusion prior with the gradient of a log-likelihood from a nonlinear model into a posterior score function described as follows:
\begin{equation}
\label{eq:detail_explain_gradient_of_poisson}
{\nabla_{\mathbf{x_0}}\mathrm{log} p(\mathbf{y}|\mathbf{{x}_0}) = -\mA^T\mathbf{y} + \mA^TD\{I_0\exp(-\mA\mathbf{{x}_0})\}},
\end{equation}
\noindent where $\{\cdot\}^\text{T}$ denotes a matrix transpose, and $\mA^\text{T}$ represents backprojection. This adaptation aligns the reconstructed images with the actual nonlinear measurements with the CT system. Efficiency in reconstruction is enhanced through a heuristic weighting scheduler. The process is further streamlined by an ordered-subset approach, leading to the order-subset DPS nonlinear (OS-DPS Nonlinear) algorithm. This innovative approach has proven both effective and efficient over existing deep learning methods.

% To alleviate the computational expense of DMs, \textcolor{red}{in the following ~\Cref{subsec:ddim}}, \citea{liu2024fourier} introduced a Fourier diffusion approach, adapted from \cite{tivnan2023}, allowing for many fewer time steps than the standard scalar diffusion model. This technique models the end-to-end process of X-ray transmission tomography and FBP to estimate image properties in sparse CT systems. They demonstrated the effectiveness of their method in reducing computational requirements while maintaining image quality on a simulated breast cone-beam CT system using sparse view acquisition.
As an example of accelerating techniques in DMs for alleviating computational burdens, discussed in  \Cref{subsec:ddim}, ~\citea{liu2024fourier} have adopted a Fourier diffusion approach originally from \cite{tivnan2023}. This method considerably cuts down the number of required time steps compared to traditional scalar DMs. Specifically tailored for X-ray transmission tomography and FBP processes, their technique efficiently estimates image properties in sparse CT systems. They validated the efficacy of this approach by demonstrating its ability to maintain image quality while reducing computational demands in a simulated breast cone-beam CT system with sparse view acquisition.

Beyond the mentioned techniques for CT reconstruction, emerging implicit neural representation (INR) methods have shown promise in addressing underdetermined CT reconstruction tasks~\cite{zang2021intratomo, reed2021dynamic, wu2023self}. However, the unsupervised nature of INR architectures offers limited constraints on the solution space, posing challenges in highly ill-posed scenarios like LA-CT and ultra sparse-view computed tomography (ultraSVCT). To bridge this gap, \citea{du2024dper} developed a hybrid network known as the Diffusion Prior Driven Neural Representation (DPER). This framework integrates a score-based diffusion model with INR optimization, employing the half quadratic splitting (HQS)~\cite{geman1995nonlinear} algorithm to enhance data fidelity. DPER leverages both the image acquisition model based on physics and the continuity prior from INR, resulting in accurate and stable CT image reconstructions. This approach delivered a superior performance on the AAPM and LIDC~\cite{armato2011lung} datasets, compared to SCOPE~\cite{wu2023self} and DiffusionMBIR~\cite{chung2023solving}.

To further alleviate the computational bottleneck in DDPMs, as investigated in \Cref{subsec:ddpm}, the patch-based DDPM, \cite{xia2022patch} was designed so that the diffusion-based inverse problem can be solved in a distributed parallel fashion.
This novel approach was demonstrated for sparse-view CT reconstruction. The network, trained on patches from fully sampled projection data, enables unsupervised inpainting of down-sampled data without the need for paired datasets. By employing a patch-based approach, this method allows for distributed parallel processing, mitigating memory issues. Additionally, a projection-domain patch-based DDPM method is inspired by using a U-Net~\cite{ronneberger2015} to learn the reverse diffusion process on fully sampled projection patches.
The reconstructed patches are finally assembled to form a high-quality projection dataset for FBP image reconstruction, Such an integration of unsupervised learning and parallel computing capability makes the diffusion model suitable for large-scale tasks. Indeed, \citea{xia2023sub} further refined the approach for breast CT reconstruction, which uses parallel DDPM in both image and sinogram domains for sparse-view reconstruction, significantly reducing radiation doses while maintaining competitive results. 

Using PFGM, as detailed in \Cref{subsec:pfgm},~\citea{ge2023jccspfgm} introduced the Joint Condition and Circle-Supervision algorithm for multiphase contrast-enhanced CT (CECT), optimizing image quality across various imaging phases. This strategy integrates temporal progression and spatial correlations to reduce the involved radiation dose.
To further enhancement of this approach, as outlined in \Cref{subsec:pfgm++}, \citea{hein2024noise} presented a PFGM++ approach to photon-counting CT (PCCT) image denoising via posterior sampling. Their strategy employs a consistency-enforced sampling scheme, using a `hijacking' technique that integrates noisy observations directly into the reverse generative process \cite{song2020denoising,chung2022come,liu2023diffusion}. This avoids complex regularization, effectively approximating the true image from noisy data \cite{karras2022elucidating}. ~\citea{hein2024poisson} further developed this framework with the Posterior Sampling Poisson Flow Consistency Model (PS-PFCM), which optimizes computational efficiency, leveraging a ``consistency distillation" process and enabling single-step or few-step function evaluations \cite{song2023consistency}.

\subsubsection{MRI Reconstruction}
Similar to deep CT image reconstruction, deep MRI image reconstruction can be viewed in two categories: model-driven and data-driven methods~\cite{geng2024dp-mdm}, although there are not clear boundary between these methods. Model-driven methods rely on mathematically derived optimization equations and incorporate artificial neural networks (ANNs) to enhance accuracy and efficiency~\cite{shlezinger2022model}. Data-driven methods extract patterns and correlations from large datasets using advanced deep learning techniques in an end-to-end fashion~\cite{arridge2019solving}. Both methods can improve MRI reconstruction, balancing accuracy, efficiency, and robustness in different ways.

\citea{chung2023decomposed} combine the Krylov subspace method~\cite{liesen2013krylov} with the diffusion model for complex imaging tasks, such as multi-coil MRI and 3D CT scans. This method, called Decomposed Diffusion Sampling (DDS), uses the fast-converging traits of the Krylov subspace method, known from traditional optimization studies, together with the strong generation abilities of the contemporary DMs.
The key to DDS is its use of the manifold constrained gradient (MCG)~\cite{chung2022improving}. DDS shows that the tangent space at a cleaned-up sample, according to Tweedie's formula, matches a Krylov subspace. This insight allows the method to perform multiple conjugate gradient (CG) updates within this space, simplifying the process significantly. 
Mathematically, this process is expressed as
\begin{equation}
    \hat{\mathbf{x}}_t' = \text{CG}(\mA^* \mA, \mA^* \mathbf{y}, \hat{\mathbf{x}}_t).
\end{equation}
Then, the DDS algorithm is defined as
\begin{equation}
    \mathbf{x}_{t-1} = \sqrt{\bar{\alpha}_{t-1}} \hat{\mathbf{x}}_t' + \tilde{\mathbf{w}}_t,
    \qquad
    \hat{\mathbf{x}}_t' = \text{CG}(\mA^*\mA, \mA^*\mathbf{y}, \hat{\mathbf{x}}_t, M), \quad M \leq l,
\end{equation}
where \(\tilde{\mathbf{w}}_t\) is the noise element, and \(M\) is the number of the CG steps within the Krylov subspace.

\citea{chung2023solving} introduced DiffusionMBIR, a novel method that combines model-based iterative reconstruction (MBIR) with 2D diffusion modeling to meet the 3D image reconstruction challenges like sparse-view tomography and compressed sensing MRI. This approach marries the precision of MBIR with the generative power of the diffusion model, enabling efficient, high-fidelity reconstructions on a single commodity GPU.
The innovation of DiffusionMBIR lies in augmenting a 2D diffusion prior with a model-based TV prior, applied specifically in the longitudinal direction at test time. This method uses the alternating direction method of multipliers (ADMM)~\cite{admm} to integrate measurement information and enforce consistency across slices:
\begin{equation}
    \begin{aligned}
        \mathbf{x}^{+} &= {\underbrace{(\mA^T\mA + \rho \mD_z^T\mD_z)}_{\mA_\text{\texttt{CG}}}}^{-1} \underbrace{(\mA^T\mathbf{y} + \rho \mD^T(\mathbf{z} - \mathbf{w}))}_{b_{\text{\texttt{CG}}}},
        \\
        \mathbf{z}^{+} &= S_{\lambda/\rho}(\mD_z \mathbf{x}^{+} + \mathbf{w}),
        \quad
        \mathbf{w}^{+} = \mathbf{w} + \mD_z \mathbf{x}^{+} - \mathbf{z}^{+}.
    \end{aligned}
\end{equation}
where ADMM updates apply soft-thresholding to the gradient of the image reconstruction (\(\mD_z \mathbf{x}^{+}\)) and adjust the dual variable (\(\mathbf{w}\)), promoting sparsity and ensuring consistency across the $z$-direction and accommodating image cubes \(256^3\) or greater.

\citea{he2024blaze3dmmarrytriplanerepresentation} also addressed high resource consumption and difficulties in capturing a natural distribution of 3D medical images, which is referred to as Blaze3DM. 
This method integrates compact triplane neural fields with a powerful diffusion model for MRI reconstruction. The method begins by optimizing data-dependent triplane embeddings and a shared decoder simultaneously to reconstruct each triplane back to the corresponding 3D volume. A lightweight 3D-aware module is used to extract the correlation among three planes. Then, a diffusion model, trained on latent triplane embeddings, achieves both unconditional and conditional triplane generation. 

During sampling, the diffusion model can be guided by degradation constraints to generate arbitrary-sized 3D volumes. Experiments demonstrate that Blaze3DM achieves the state-of-the-art performance and improves computational efficiency compared to existing 3D medical inverse problem solvers, including compressed-sensing MRI, DiffusionMBIR~\cite{chung2023solving} and TPDM~\cite{lee2023improving}.

To further enhance MRI quality,~\citea{geng2024dp-mdm} introduced a comprehensive detail-preserving method using multiple DMs to extract structural and textural features in the k-space. Traditional DMs often struggle with capturing intricate details accurately~\cite{shin2014calibrationless, haldar2016p}. The proposed method employs virtual binary modal masks to refine the value range of k-space data, which places the model focus within a highly adaptive center window. An inverted pyramid structure is used, allowing for a multi-scale representation. Empirically, a step-by-step refinement significantly enhances detailed image features on the selected benchmark datasets.

Similarly, \citea{guan2024correlated} proposed another MRI reconstruction scheme that leverages k-space data properties and the diffusion process to enhance image quality. By mining multi-frequency priors using diverse strategies, the proposed method preserves image texture and accelerates the sampling process. A key to this method is the Correlated and Multi-frequency Diffusion Modeling (CM-DM), which combines the \textit{Weight-K-Space}' and \textit{Mask-K-Space}' strategies to optimize the utilization of hidden information. This scheme outperforms the state-of-the-art methods in maintaining image details, not only improving reconstruction accuracy but also facilitating fast convergence by focusing on high-frequency priors.

\subsubsection{Nuclear Imaging}

In the previous sub-sections, we comprehensively discussed the applications of physics-inspired generative models in CT and MRI reconstruction tasks. In this sub-section, we want to show that physics-informed generative models can be also utilized for other imaging modalities with an emphasis on PET image reconstruction. 

PET image reconstruction faces challenges such as high variance Poisson noise and a wide dynamic range. To address these issues, \citea{Singh_2024} proposed PET-specific score-based generative models, including a framework for both 2D and 3D PET and an extension for guided reconstruction using MRI images. Validated through experiments on patient-realistic data, the proposed method demonstrates robust results with potential for improving PET reconstruction quality.

As another innovative approach for PET imaging,~\citea{chen202425dmultiview} introduced the 2.5D Multi-view Averaging Diffusion Model (MADM). This model converts non-attenuation-corrected low-dose PET (NAC-LDPET) into attenuation-corrected standard-dose PET (AC-SDPET), lowering the tracer dose and eliminating the need for CT-based attenuation correction. MADM uses separate DMs for axial, coronal, and sagittal views, averaging outputs to produce high-quality 3D images. The 2.5D architecture reduces computational demands and mitigates 2D discontinuities. In their experiments, MADM outperformed the competitive methods, offering high-quality 3D PET images with reduced radiation exposure and improved computational efficiency.

To further address the challenges of generalizability across different image noise levels, acquisition protocols, and patient populations in low-dose PET imaging, \citea{xie2024dose} developed DDPET-3D, a dose-aware diffusion model specifically for 3D low-dose PET imaging. Traditional DMs struggled with consistent 3D reconstructions and generalization across varying noise levels, often resulting in visually appealing but distorted images. DDPET-3D was extensively evaluated using 9,783 F-FDG studies from 1,596 patients across four medical centers with different scanners and clinical protocols. The model demonstrated strong generalizability to different low-dose levels, scanners, and protocols. Reader studies by nuclear medicine physicians confirmed that DDPET-3D produced superior denoised results comparable to or better than 100\% full-count images and previous deep learning baselines. Real low-dose scans were included to showcase the clinical potential of DDPET-3D, indicating the feasibility of achieving high-quality low-dose PET images.

Given the limited space of this review, we will not delve into details on the applications of the diffusion model in other imaging modalities such as ultrasound imaging~\cite{zhang2023ultrasound, zhang2024ultrasound}, histopathology~\cite{sridhar2023patho}, and endoscopy~\cite{chen2024lightdiff}. 

\subsection{Medical Image Generation}

Generating synthetic images through advanced generative models is a crucial area that addresses the scarcity of annotated medical datasets and ethical concerns associated with using real patient data. Techniques based on physics-informed models and their derivatives, explored in \Cref{sec:phys-gms}, are particularly effective at producing high-quality, diverse synthetic images that accurately replicate human anatomy and disease variations, providing valuable resources for training algorithms and testing new technologies~\cite{cvpr2017}.

\subsubsection{High-Quality Image Synthesis}

DMs have revolutionized medical image synthesis by addressing dataset scarcity, memory constraints, and computational complexity. These models generate high-quality images realistically, providing essential resources for developing and testing medical imaging algorithms~\cite{sherwani2024systematic}.

To address the limited availability of 3D medical image datasets, \citea{zhu2024generative} developed GEM-3D, a generative approach using conditional DMs to synthesize and enhance 3D medical images. GEM-3D begins with a 2D slice containing patient-specific information and uses a 3D segmentation mask to guide the generation process. By decomposing 3D images into masks and patient prior information, GEM-3D enables versatile image generation, counterfactual synthesis, and dataset enhancement. This approach improves the quality and variability of generated 3D images, tackling high-quality 3D generation and condition decoupling for re-sampling data. The key innovation lies in decomposing segmentation masks and patient information for data augmentation, improving distribution coverage.

Furthermore, \citea{ju2024vmddpm} introduced a hybrid diffusion model based on a State Space Model (SSM). 
This model combines the local perception capabilities of CNNs with the global attention of SSM, while maintaining linear computational complexity~\cite{heidari2024computationefficient}. The architecture features a multi-level feature extraction module called the Multi-level State Space Block (MSSBlock) and an encoder-decoder structure with the State Space Layer (SSLayer), designed specifically for medical pathological images. Additionally, a Sequence Regeneration strategy improves the cross-scan module (CSM), enabling the Mamba~\cite{gu2024mamba} module to fully capture image spatial features and enhance model generalization. This hybrid architecture produced a competitive performance, according to a qualitative assessment by radiologists.

To ease the tension between high-dimensional data and limited GPU memory~\cite{zhu2023makeavolume}, ~\cite{friedrich2024wdm} introduced wavelet diffusion models (WDMs), a wavelet-based medical image synthesis framework that applies DMs to wavelet-decomposed images. WDM generates high-resolution 3D medical images by creating synthetic wavelet coefficients and then applying inverse discrete wavelet transform (IDWT). By decomposing an input image into a limited number of wavelet coefficients, concatenated into a single target matrix for the diffusion model, WDM maintains the standard architecture's width through an initial convolution. Operating in the wavelet domain reduces spatial dimensions significantly, allowing for shallower networks, fewer computations, and reduced memory usage. During training, a corrupted sample is drawn, and the diffusion model approximates the original wavelet coefficients in terms of mean squared error. Inference starts with a sample from a normal distribution, and goes through reverse diffusion to reconstruct a final image. In the experiments, WDM outperformed the state-of-the-art methods in high-resolution 3D medical image generation.

In another study, \citea{huang2024memory} introduced a cascaded amortized latent diffusion model (CA-LDM) to  synthesize high-resolution OCT volumes efficiently. This model employs non-holistic auto-encoders (NHAE) to map between low-resolution latent spaces, as mentioned in \Cref{subsec:ldm}, and high-resolution volumes, achieving volumetric super-resolution slice-by-slice to manage the memory constraint. Cascaded diffusion processes further enhance this memory efficiency by first synthesizing the 3D global latent representation, followed by a 2D slice-wise refinement for high-resolution details. This approach effectively balances memory and speed in a task-specific way.

Yet another approach to reducing the number of inference steps is Fast-DDPM~\cite{jiang2024fast}. This model reduces training and sampling to just 10 time steps, optimizing the process with an appropriate noise scheduler. By aligning training and sampling procedures, Fast-DDPM enhances training speed, sampling speed, and generation quality. Evaluated on important tasks such as multi-image super-resolution, image denoising, and image-to-image translation, Fast-DDPM demonstrated its effectiveness in minimizing computational overhead while maintaining high performance relevant to medical imaging applications.

\subsubsection{Image Segmentation and Anatomical Guidance}

In medical image generation, leveraging anatomical structures significantly enhances the precision and robustness of synthesized images. This approach targets the inherent complexity and variability of human anatomy, ensuring that generated images are clinically relevant. By incorporating anatomical guidance, these techniques provide major improvements in the fidelity and utility of synthesized medical images, which are essential for training and validating deep imaging algorithms.

To further enhance medical image generation, \citea{konz2024generation} proposed SegGuidedDiff, a diffusion model using multi-class anatomical segmentation masks at each sampling step. This method incorporates a random mask ablation training stage, allowing the model to condition on specific anatomical constraints while maintaining flexibility. SegGuidedDiff generates images directly from anatomical segmentation maps, ensuring precise spatial guidance. The evaluative studies on breast MRI and abdominal/neck-to-pelvis CT datasets demonstrated that SegGuidedDiff achieved excellent performance, producing realistic images faithful to the input masks.

Similarly, LeFusion, a lesion-focused diffusion model~\cite{zhang2024lefusion}, addresses long-tail imbalance and algorithmic unfairness in clinical data by focusing on lesion areas~\cite{chen2022algorithm}. Inspired by diffusion-based inpainting, LeFusion integrates the background context into the reverse diffusion process, preserving background quality while handling multi-class lesions flexibly. This method diversifies image synthesis by introducing lesion masks. On the DE-MRI cardiac lesion segmentation dataset (Emidec)~\cite{lalande2021deep}, LeFusion significantly outperformed the nnUNet model's performance. 

\subsubsection{Data Imbalance and Image Robustness}
Addressing data imbalance and enhancing model robustness are critical in medical image generation. These issues often induce biases and reduce generalizability. 

\citea{chambon2022roentgen} introduced a vision-language foundation model called RoentGen to generate synthetic chest X-rays (CXR) based on text prompts. It leverages a pre-trained LDM fine-tuned on publicly available CXRs and radiology reports for the creation of diverse, high-quality images guided by radiology-specific text prompts. This method enhances image classification tasks through data augmentation. RoentGen's architecture includes a VAE, a conditional denoising U-Net, and a CLIP text encoder. The U-Net and text encoder are fine-tuned on the MIMIC-CXR dataset while the VAE is kept frozen. The model samples noise from a Gaussian distribution and predicts the original noise using the U-Net and encoded text prompts. 

Another contribution to this field is Cheff, a cascaded latent diffusion model introduced by \citea{weber2023cascaded}. Cheff generates highly realistic 1-megapixel chest radiographs and addresses class imbalances by leveraging ``MaCheX", the largest open collection of chest X-ray datasets, which supports report-to-chest-X-ray generation \cite{sundaram2021ganbased}. This methodology integrates auto-encoders and super-resolution DMs, setting a new standard in radiological image synthesis.

Additionally, the MedM2G framework \cite{zhan2024medm2g} presents a unified approach for medical multi-modal generation. MedM2G overcomes the limitations of separate generative models by aligning and generating medical data across multiple modalities (CT, MRI, X-ray) within a single model. It makes a central alignment to integrate diverse medical datasets, preserves specific clinical knowledge through visual invariability, and enhances cross-modal interactions using adaptive diffusion parameters. MedM2G outperforms the state-of-the-art models in the relevant tasks including text-to-image and image-to-text generation, in terms of accuracy and efficiency across multiple medical datasets.

\subsubsection{Innovative Frameworks for Image Generation}
Developing innovative frameworks for generative AI is essential for advancing the quality and utility of synthetic images. This includes methodologies for inpainting, counterfactual generation, and multi-modal alignment, which will be instrumental in meeting several medical imaging challenges. 

The Concept Discovery through Latent diffusion-based counterfactual trajectories (CDCT) framework \cite{varshney2024generating} addresses classifier biases and uncovers medical concepts by using LDMs to generate counterfactual trajectories. This approach involves generating trajectories with LDMs, disentangling semantic spaces with a VAE, and identifying relevant concepts through a search algorithm. This method significantly improves high-fidelity, patient-specific 3D medical image generation.

Another novel technique utilizes voxel-wise noise scheduling in LDM for inpainting pathological features into lumbar spine MRIs \cite{hansen2024inpainting}. This method overcomes the limitations of traditional data augmentation models by ensuring better anatomical integration, significantly improving the results of lumbar spine MRI.

\subsection{Medical Image Analysis}

In medical image analysis, innovative techniques are continually emerging in various clinical tasks. Recent advancements in texture restoration~\cite{vazia2024diffusion, wang2024implicit, xu2024maediff}, denoising~\cite{hein2024noise, xiong2024qsmdiff}, anomaly detection~\cite{gao2024u2mrpd, siddiqui2024valdmd, du2023boosting, bercea2024diffusion}, and segmentation~\cite{chen2023berdiff, wu2023medsegdiffv2, zhang2024cts, bogensperger2024flowsdf} have significantly improved the quality of medical images, suggesting a great potential of contemporary AI generative models.  A few representative examples are as follows.

\citea{wang2024implicit} introduced the implicit image-to-image Schrödinger bridge (I\(^3\)SB), an enhanced version of the Image-to-Image Schrödinger Bridge (I\(^2\)SB)~\cite{liu2023i2sb}. I\(^3\)SB employs a non-Markovian process that integrates corrupted images at each generative step, significantly improving texture restoration and accelerating generation, particularly in CT super-resolution and denoising tasks. By establishing a diffusion bridge between clean and corrupted images, I\(^3\)SB optimizes generative steps to produce high-quality images. This method outperformed traditional models like cDDPM on the RPLHR-CT-tiny~\cite{yu2022rplhr} datasets. 

\citea{Chen_2023berdiff} designed the Bernoulli Diffusion model for medical image segmentation (BerDiff), which improves traditional diffusion methods by incorporating Bernoulli noise, making it better suited for binary segmentation tasks~\cite{amit2022segdiff, wu2023medsegdiff}. This approach generates more accurate and diverse segmentation masks, crucial for delineating ambiguous tumor boundaries. BerDiff provides multiple plausible segmentation options by sampling different initial noises and intermediate states. Evaluated on various medical imaging datasets, BerDiff outperformed the current methods, showcasing its potential for clinical imaging tasks~\cite{wolleb2021diffusion}.

To shorten prediction time of the diffusion model,~\citea{zhang2024cts} presented CTS, a consistency-based medical image segmentation model, as noted in \Cref{subsec:ddim}. By condensing sampling into a single iteration, CTS accelerates training and prediction while maintaining the generation quality~\cite{song2023consistency}. This model uses multi-scale feature supervision and customized loss functions to improve segmentation and convergence. It integrates multi-scale signals into the UNet model and iteratively refines parameters, capitalizing contextual information. 

In response to the limitations of 2D methods, which often miss crucial out-of-slice information,~\citea{kim2024adaptive}. introduced a refined LDM with a novel switchable block, the multiple switchable spatially adaptive normalization (MS-SPADE). This innovation enables effective 3D medical image translation without patch cropping, allowing for the translation of multiple target modalities from a single source. The MS-SPADE block dynamically adjusts source latents to match target styles, improving fidelity and reducing the need for multiple models, thereby addressing practical challenges like high scan costs and limited availability in clinical settings.

Furthermore, \citea{arslan2024selfconsistent} address key challenges in DDMs used for medical image translation, particularly the weak guidance from source modalities and the poor alignment between denoising and the necessary transformations from source to target that reduce effectiveness \cite{gungor2023adaptive}. They present the Self-consistent Recursive Diffusion Bridge (SelfRDB). This model markedly betters traditional DDMs by introducing a forward process with a strategy that gradually increases noise levels, improving generalization and boosting the transfer of information between modalities. SelfRDB stands out from typical DMs by using a special recursive estimation method in its backward steps, which permits ongoing improvement of the target image while keeping the original tissue details intact. Extensive tests on multi-contrast MRI and MRI-CT translations show that SelfRDB surpasses current GAN~\cite{zhou2020hi} and DMs and is the first to employ a soft-prior on the source modality and a consistent recursive estimation to refine the accuracy of backward diffusion steps.

\citea{zheng2024deformation} presented the Deformation-Recovery Diffusion Model (DRDM), a novel approach that shifts from traditional intensity-based synthesis~\cite{song2020denoising} to deformation-based methods~\cite{deformation-based}, using a mechanism that generates topologically-preserving multi-scale Deformation Velocity Fields (DVFs). This model enhances the interpretability and anatomical accuracy of synthetic medical images by creating physically plausible deformation fields. Capable of managing instance-specific deformations without reference images, the DRDM supports applications such as data augmentation for few-shot learning and synthetic training for image registration. This advancement enriches synthetic medical image generation and aids critical downstream tasks without the need for external annotations.

Finally, FlowSDF~\cite{bogensperger2024flowsdf} is another image-guided conditional flow matching framework through the signed distance function (SDF), facilitating an implicit distribution of segmentation masks. By leveraging SDF, FlowSDF achieves more natural distortions and high-quality sampling from the segmentation mask distribution of interest, enhancing predictive robustness with statistical tools like uncertainty maps. This method employs ODE-based flow matching~\cite{lipman2022flow, liu2022flow} to model probability flow, ensuring a smoother distribution and natural transitions between class modes. FlowSDF integrates image-guided flow matching with SDFs, enabling noise-injection for generating multiple segmentation maps and uncertainty quantification, thus improving model robustness and interpretability.

    \section{FUTURE DIRECTIONS}

Physics-inspired GMs excel at closely matching the distribution of real images. This has not only paved the way for various medical imaging applications (as outlined in the previous section), but also offers potential for future developments along several research directions. Since hot topics for DDPMs are comprehensively covered in~\cite{kazerouni2023}, this section focuses on other types of physics-inspired GMs, as well as more general aspects such as unification of GMs and performance considerations.

%--------------------------------------------------------
\subsection{Unification of SDMs}
%--------------------------------------------------------

DDPMs (cf. Section 2.1) are grounded on the concept of non-equilibrium thermodynamics, specifically employing a diffusion-like process to progressively shape random noise into coherent data. 
PFGMs (cf. Section 2.4) are based on electrostatics, where the Poisson equation is used to describe the evolution of particles under the influence of electrostatic force.
Accordingly, the question arises whether other physical phenomena can also give rise to GMs, or in other words, whether a duality between a physical phenomenon and a respective ``score-matching'' GM exists~\cite{liu2023}. ~\citea{liu2023} show that this can indeed be the case, under several restrictions on the physical processes. This opens up a wide range of new options that take advantage of physical processes for advancing medical imaging research.

%--------------------------------------------------------
\subsection{Integration with VLMs}
%--------------------------------------------------------

Whilst DMs have shown excellent performance on various image generation tasks(both in medical imaging as detailed in Section 3, as well as various other fields), Large Language Models (LLMs) are dominant for generative tasks in natural language processing (NLP). Nevertheless, if a dedicated visual tokenizer that maps pixel-space inputs to discrete tokens can be effectively analyzed by LLMs, the resultant vision-language models may even outperform DMs on image and video generation benchmarks including ImageNet and Kinetics~\cite{yu2024language}.

Instead of using either a DM or a LLM, an increasing number of research efforts are capitalizing on both types of models synergistically. The canonical way of obtaining VLMs, or more generally multimodal models, would be to combine several models; for example, an image encoder to understand images, an LLM to generate text, and a DM to generate images. A very interesting recent development is Chameleon~\cite{team2024chameleon}, which represents a new approach of early-fusion token-based mixed-modal models designed to comprehend and produce images and text in any order. 
The flexibility of such models may prove useful in the field of medical imaging. 

Integrating DMs and LLMs offers the following potential benefits. \Romannum{1}) Diverse strengths: Each model type excels in a unique aspect; combining them leverages their strengths. \Romannum{2}) Complementary information: Language models extract semantic information from text prompts, guiding the DM for more accurate results (e.g. in image generation or image reconstruction). \Romannum{3}) Progressive refinement: LLMs provide initial concepts, which may be refined by score-matching models into more realistic images, yielding higher fidelity. \Romannum{4}) Task simplification: Integrating models specializing in different sub-tasks divides and conquers the complexity of tasks, thus enhancing efficiency, which is highly desirable for many challenging medical imaging applications.

However, in spite of recent advancements in text-to-image generation with diffusion or score-matching models, they often struggle to accurately follow the prompts when spatial or common sense reasoning is required~\cite{lian2024llmgrounded}. Vice versa, adaptation of LLMs to medical imaging is not straight-forward, since they have safeguards regarding medical imaging data. As a result, performance of proprietary LLMs, such as ChatGPT-4, is usually quite poor for medical image synthesis and thus domain adaptation using task-specific medical datasets is often required. 

Therefore, extensive research efforts are under way to explore how to link both models and how to take advantage of the strengths of each model for generative tasks. In this context, there are several studies on different types of GMs. For instance, in~\cite{lian2024llmgrounded}, a DM for text-to-image scene generation is enhanced with a LLM that generates a scene layout with captioned bounding boxes to handle more complex prompts. In~\cite{chen2024lowdose} an autoencoder is enhanced with high-level features from a pre-trained LLM and its vocabulary. Regarding PFGMs and PFGM++ (cf. Sections 2.4 and 2.5), their integration with LLMs is largely untouched so far. Therefore, further research is clearly needed.

%--------------------------------------------------------
\subsection{Novel Applications of GMs}
%--------------------------------------------------------

Physics-inspired GMs have shown great potential in the context of various medical imaging modalities (cf. Section 3.1) and associated tasks (cf. Sections 3.2 and 3.3). It is expected that these models will be adapted to novel imaging modalities, such as phase-contrast and dark-field imaging techniques, spectral imaging, or multi-modal and multi-contrast approaches.
Furthermore, generalist DMs are attractive, because these foundation models will be able to simultaneously cover multiple medical image domains and handle diverse clinical tasks.
By taking advantage of multiple domains simultaneously, more robust performance on clinical tasks with scarce data is to be expected using the AI generalist approach.

Despite these potentials, there are several limitations regarding the score-matching generative models in medical imaging. Medical image data is always linked with protected health information (PHI), and protecting PHI is a critical aspect that needs to be taken care of. Although some of the proposed DMs could have a large clinical impact in medical imaging and image analysis, in many cases limitations arise due to the lack of data availability limited by the PHI concerns. This is particularly evident in applications such as personalized medicine, rare diseases, longitudinal studies, and multi-institutional collaborations. Moreover, different imaging modalities, varying quality of scans and diverse patient demographics are inherent characteristics of medical image data. Furthermore, insufficient computing resources is an obstacle for DM training when addressing non-trivial datasets, given an extensive computational cost of DMs in general. Memory efficient methods, such as Low-Rank Adaptation (LoRA)~\cite{hu2021}, offer some democratization of the field by enabling users to adapt DMs to their domain of interest at a lower computational cost. Finally, general theoretical insights regarding score-matching GMs would be generally desirable and particularly in medical imaging areas. Further theoretical insights and practical solutions are anticipated to address potential biases, instability, hallucination, and sub-optimality in training quality and efficiency.
    
    % conclusion
    \section{CONCLUSION}

In this article, we have provided a general overview of different physics-inspired GMs and associated acceleration techniques. Furthermore, we have highlighted their applications in different medical imaging scenarios, including medical image reconstruction, image generation, and image analysis. Finally, we have discussed future research directions, such as unification of GMs, integration with VLMs, new applications of GMs, including foundation models, as well as challenges to overcome. Clearly, there is immense potential for physics-inspired GMs and great opportunities to solve the current problems and redefine medical imaging. 
    
    % <<<<<<<<<<<<<<<<<<<<<<<<<<<<<<<<<<<<<<<<<<<
    
    %%Harvard
    % \bibliographystyle{unsrtnat}
    % \bibliographystyle{unsrt}
    \bibliographystyle{plainnat}
    \bibliography{ref}

    % appendices
    %\newpage
    %\input{content/appendices/PFGMpp}

\end{document}